\documentclass[journal]{vgtc}                
\ifpdf
  \pdfoutput=1\relax                   
  \usepackage{graphicx}                
  \DeclareGraphicsExtensions{.pdf,.png,.jpg,.jpeg} 
\else
  \usepackage{graphicx}                
  \DeclareGraphicsExtensions{.eps}     
\fi%

\graphicspath{{figures/}{pictures/}{images/}{./}} 

\usepackage{microtype}                 
\PassOptionsToPackage{warn}{textcomp}  
\usepackage{textcomp}                  
\usepackage{mathptmx}                  
\usepackage{times}                     
\usepackage{cite}                      
\usepackage{tabu}                      
\usepackage{booktabs}                  

\usepackage{enumitem}
\setitemize{noitemsep,topsep=5pt,parsep=0pt,partopsep=0pt}
\usepackage{xspace}
\usepackage{xstring}
\usepackage[normalem]{ulem} 
\usepackage{amsmath}
\usepackage{bbm}
\usepackage{amssymb}
\usepackage[mathscr]{euscript}
\usepackage{algorithm}
\usepackage[noend]{algpseudocode}

\newcommand{\note}[2]{\textcolor{#1}{}} 

\newcommand{\breakline}{\vskip \medskipamount}
\newcommand{\q}[1]{``\textit{#1}''}
\newcommand{\fakeTitle}[1]{\breakline \noindent \textbf{#1}.}

\newcommand{\tim}[1]{\note{magenta}{TA: #1}}
\newcommand{\vv}[1]{\textcolor{black}{#1}}

\newcommand{\ie}{{\em i.e.,}\xspace}
\newcommand{\eg}{{\em e.g.,}\xspace}

\newcommand{\etal}{{\em et~al.}\xspace}

\newcommand{\apriori}{{\em a priori}\xspace}

\newcommand{\R}{\mathbb{R}}
\newcommand{\D}{\mathscr{D}}

\newcommand{\figureTeaser}{
\teaser{
  \vspace{0pt}
  \centering
  	\includegraphics[width=\linewidth,trim={0cm 8cm 1cm 0cm}, clip]{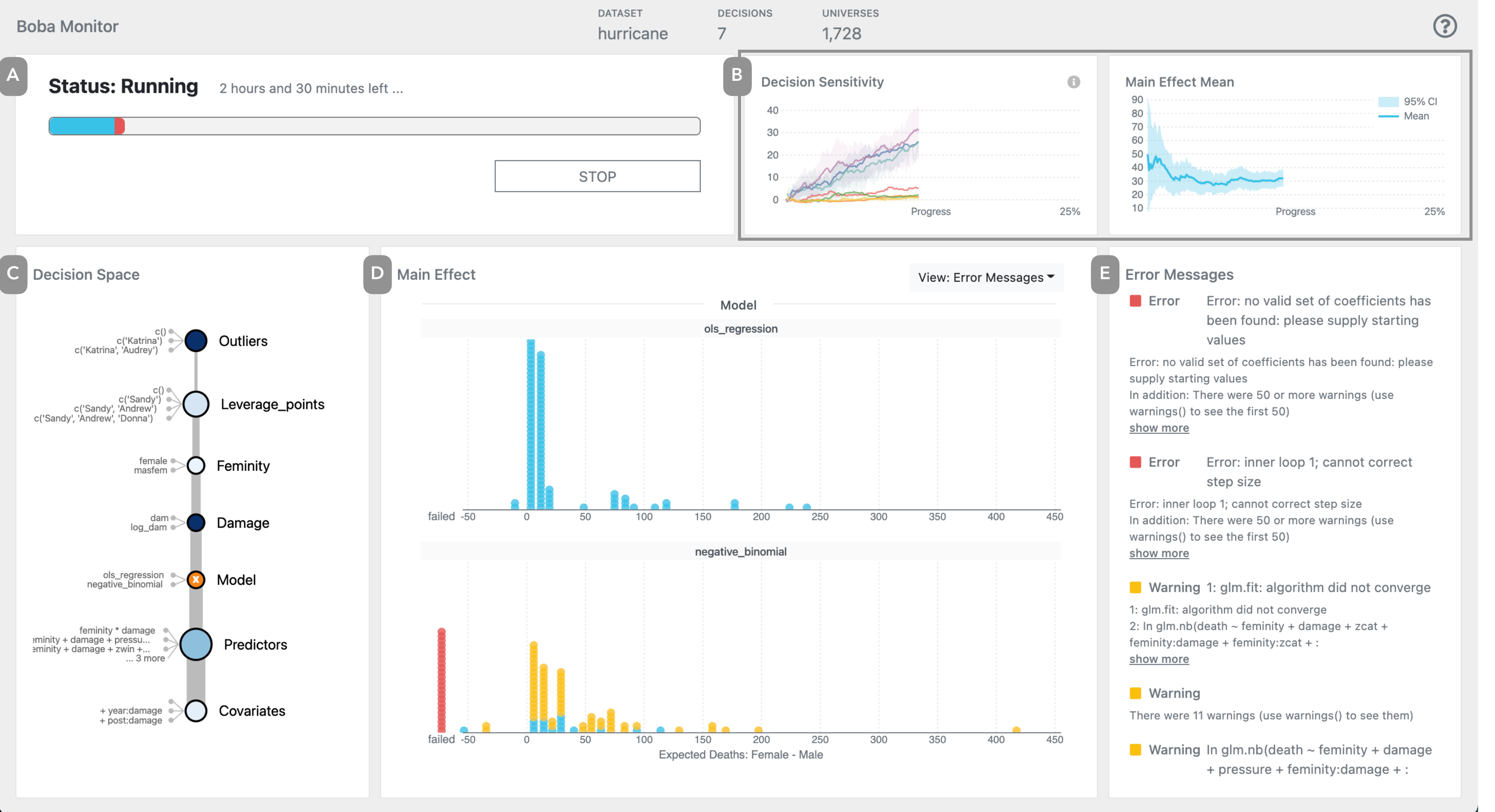}
  \vspace{-20pt}
  \caption{Leveraging underlying approximation algorithms, the Boba Monitor enables analysts to monitor progress and diagnose issues while a multiverse analysis is running. Analysts can control the execution on the fly (a), observe progressive estimates of decision sensitivity and effect size (b), and gauge when the approximation has achieved reasonable convergence. To assess validity, analysts may reflect on the decision space structure (c), examine the range of effect size estimates (d), and review runtime errors and warnings (e). Clicking a decision node (c) allows users to compare between options and identify which option(s) lead to specific issues (d).}
	\label{fig:teaser}
}
}

\newcommand{\figureResultBox}{
\begin{figure*}[t]
	\vspace{-15pt}
	\centering
	\includegraphics[width=\linewidth,trim={0.5cm 25cm 0cm 0cm}, clip]{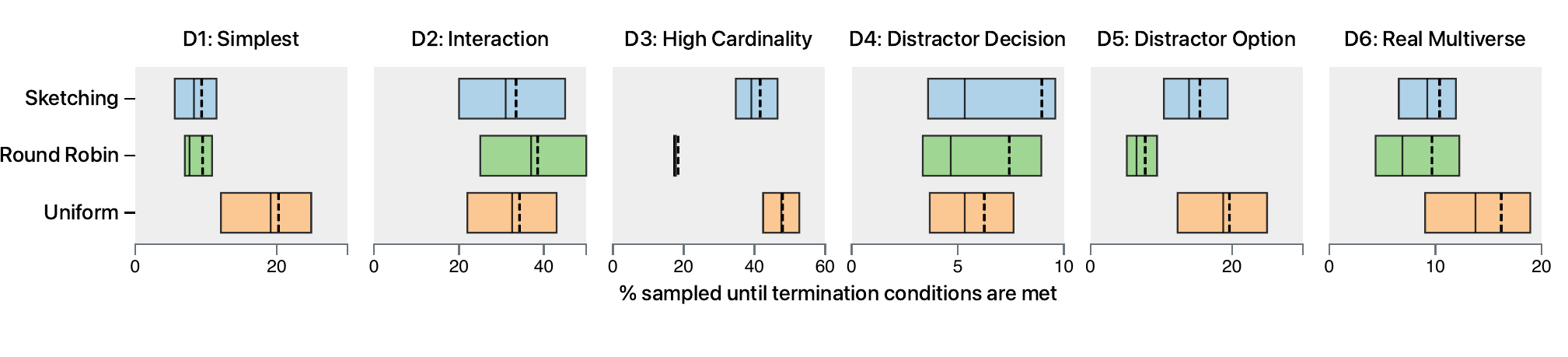}
	\vspace{-36pt}
	\caption{Empirical evaluation of the approximation algorithms on six datasets. The x-axis indicates the percentage of the full multiverse sampled until the algorithms accurately estimate and rank sensitive decisions (lower is better). The box plot shows the median and IQR across 200 runs using different random seeds, and the dashed line represents the mean.}
	\label{fig:result-box}
	\vspace{-16pt}
\end{figure*}
}

\newcommand{\figureResultBias}{
\begin{figure}
	\centering
	\includegraphics[width=1.0\columnwidth,trim={0cm 25.5cm 9cm 0cm}, clip]{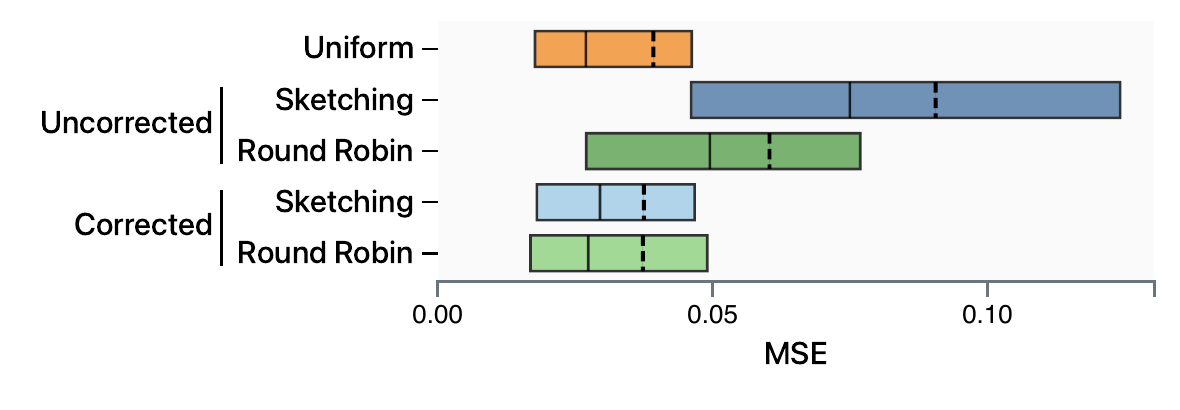}
	\vspace{-28pt}
	\caption{Empirical evaluation of bias correction in mean estimation. The x-axis shows the MSE between the estimated and actual mean (lower is better). The box plot shows the median and IQR across 200 runs, and the dashed line represents the mean. }
	\label{fig:result-bias}
	\vspace{-13pt}
\end{figure}
}

\newcommand{\figureResultPearson}{
\begin{figure*}[t]
\vspace{-15pt}
	\centering
	\includegraphics[width=\linewidth,trim={0cm 25cm 0cm 0cm}, clip]{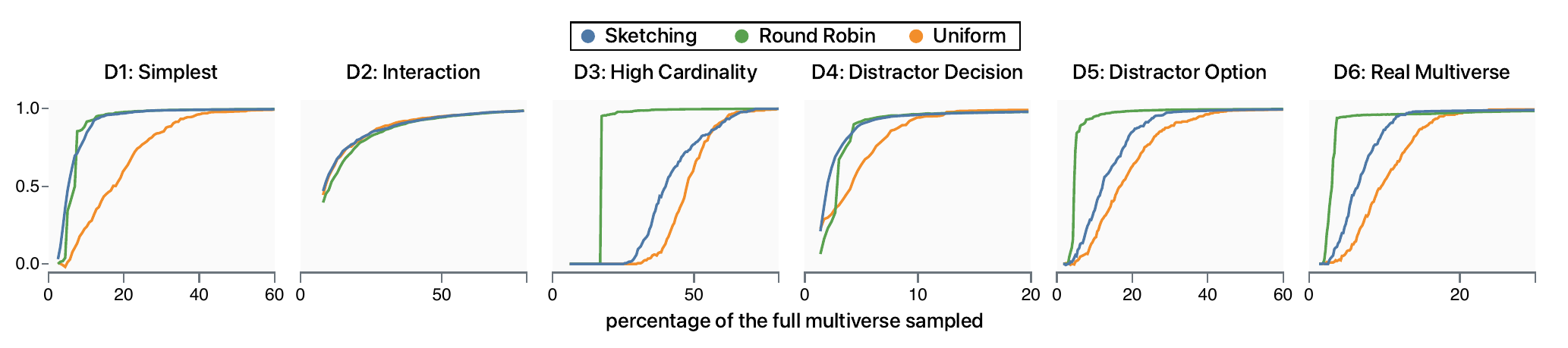}
	\vspace{-24pt}
	\caption{Pearson correlation between sample sensitivity and the ground truth over time. The x-axis encodes the percentage of the full multiverse sampled (lower is better) and the y-axis encodes the correlation coefficient (higher is better).}
	\label{fig:result-pearson}
	\vspace{-16pt}
\end{figure*}
}

\newcommand{\figureResultSpearman}{
\begin{figure}
	\centering
	\includegraphics[width=1.0\columnwidth,,trim={0cm 25cm 10.2cm 0.7cm}, clip]{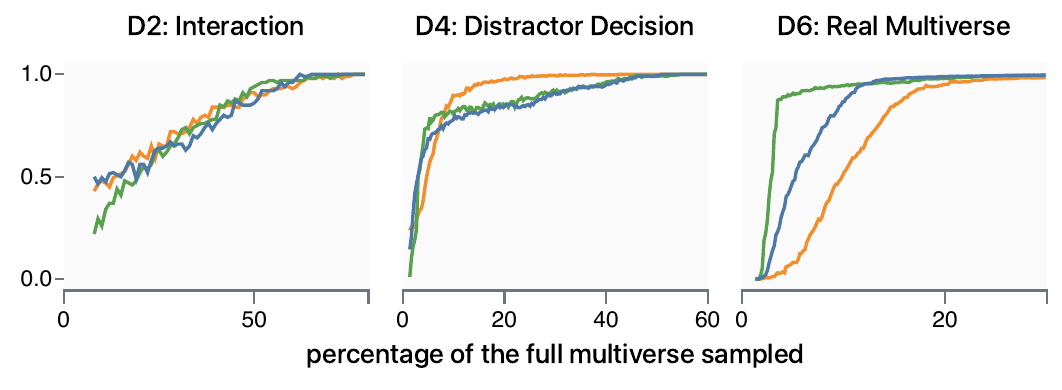}
	\vspace{-24pt}
	\caption{Spearman's rank correlation between sample sensitivity and the ground truth over time. The x-axis encodes the percentage of the full multiverse sampled and the y-axis encodes the correlation coefficient. Datasets with only one sensitive decision are omitted.}
	\label{fig:result-spearman}
	\vspace{-13pt}
\end{figure}
}

\newcommand{\figureMortgage}{
\begin{figure}[t]
	\centering
	\includegraphics[width=1.0\columnwidth,trim={0cm 9cm 16cm 0cm}, clip]{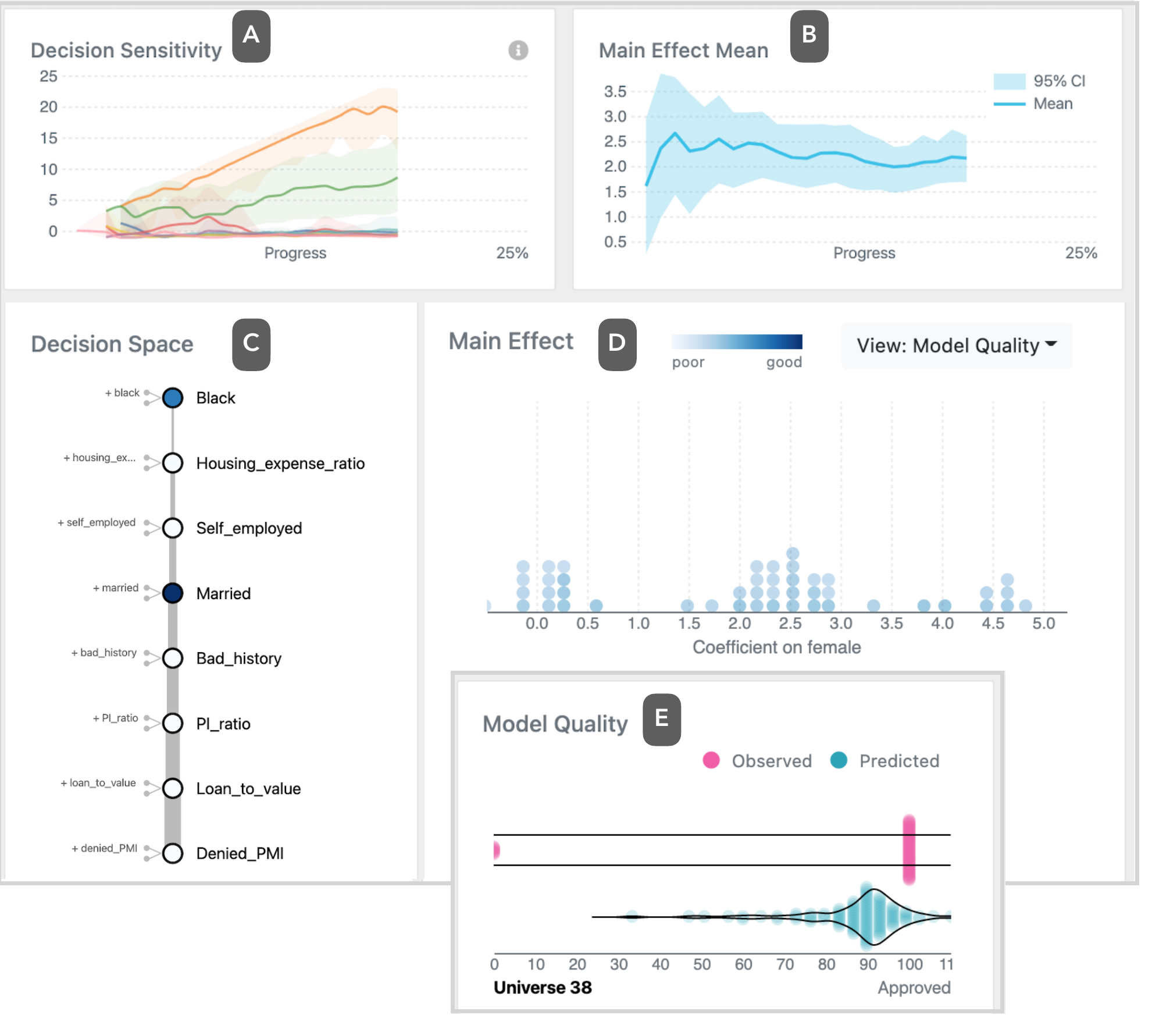}
	\vspace{-24pt}
	\caption{A case study on mortgage analysis. With less than one fifth of the multiverse completed, the sampling estimates converge reasonably: sensitive decisions are distinct from non-sensitive ones (a), and the mean effect size estimate remains stable (b). The two sensitive decisions (c), \textit{black} and \textit{married}, agree with prior work~\cite{young2017}. However, model quality checks show an unsatisfactory quality throughout (d) and indicate a large mismatch between observed and predicted data (e).}
	\label{fig:mortgage}
	\vspace{-10pt}
\end{figure}
}

\newcommand{\figureCaseTwo}{
\begin{figure*}[t]
\vspace{-10pt}
	\centering
	\includegraphics[width=1\linewidth,trim={0cm 16.5cm 0cm 0cm}, clip]{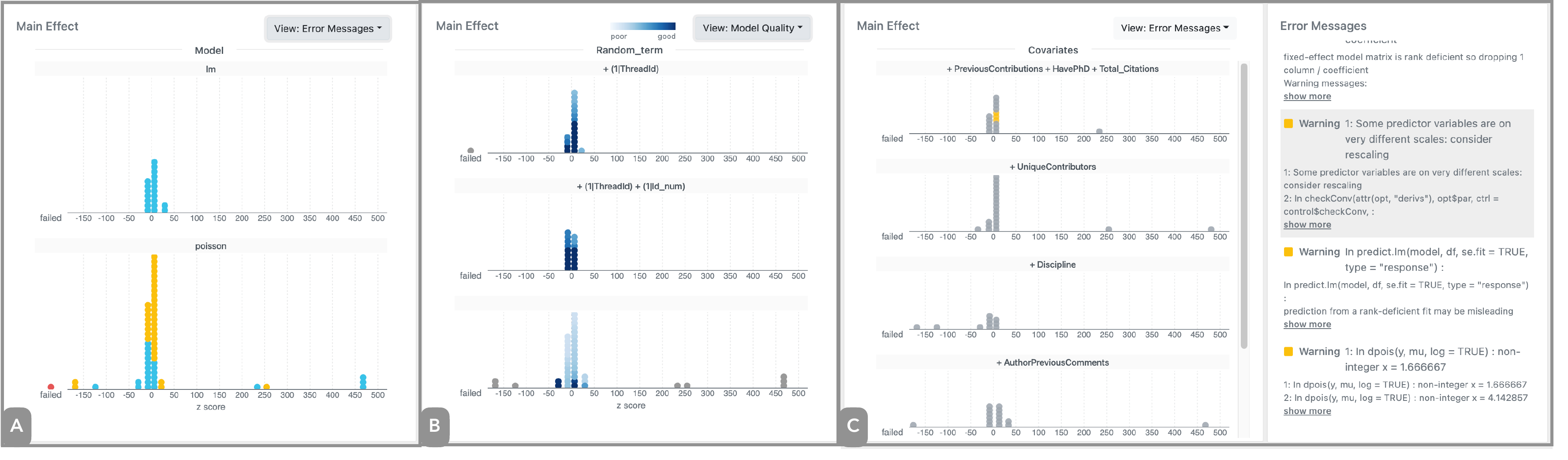}
	\vspace{-18pt}
	\caption{A case study illustrating error diagnostics of a multiverse on scientific debate. (a) The Poisson regression model is responsible for all errors, warnings, and abnormal point estimates. (b) Adding random effects greatly improves model quality and removes outlier estimates. (c) A warning of predictors having different scales only occurs in one set of covariates, suggesting the fix of rescaling these covariates.}
	\label{fig:case}
	\vspace{-18pt}
\end{figure*}
}

\newcommand{\figureModelQuality}{
\begin{figure}[t]
	\centering
	\includegraphics[width=1.0\columnwidth, trim={0cm 13cm 14cm 0cm}, clip]{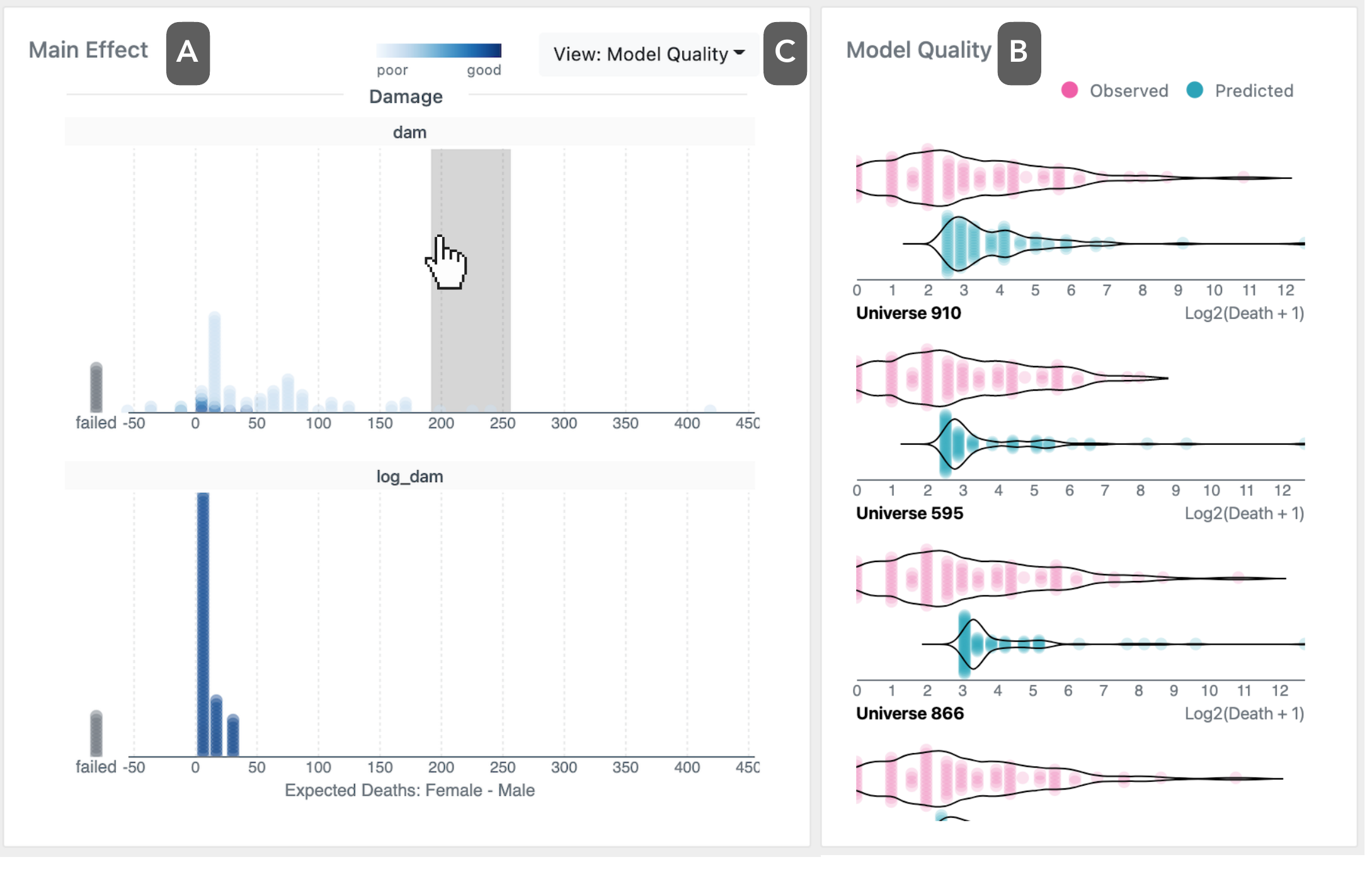}
	\vspace{-22pt}
    \caption{To enable an overview of model quality, universes are colored according to a quantitative model quality metric, with a lighter blue indicating a poorer fit (a). Brushing the main effect view populates the model quality view (b) with visual predictive checks that compare predicted data (blue) with observed data (pink). Users toggle between viewing error information and model quality using a dropdown menu (c).}
    \label{fig:system-model-quality}
    \vspace{-13pt}
\end{figure}
}

\onlineid{0}

\vgtccategory{Research}
\vgtcpapertype{Area 5 -- Data Transformation}

\title{Approximation and Progressive Display of Multiverse Analyses}

\author{Yang Liu, Tim Althoff, and Jeffrey Heer}
\authorfooter{
\item
 All authors are with the University of Washington. E-mails: yliu0, althoff, jheer@uw.edu.
}

\shortauthortitle{Liu \MakeLowercase{\textit{et al.}}: Multiverse Sampling}

\abstract{A multiverse analysis evaluates all combinations of ``reasonable’’ analytic decisions to promote robustness and transparency, but can lead to a combinatorial explosion of analyses to compute.
Long delays before assessing results prevent users from diagnosing errors and iterating early.
We contribute (1) \emph{approximation algorithms} for estimating multiverse sensitivity and (2) \emph{monitoring visualizations} for assessing progress and controlling execution on the fly.
We evaluate how quickly three sampling-based algorithms converge to accurately rank sensitive decisions in both synthetic and real multiverse analyses.
Compared to uniform random sampling, round robin and sketching approaches are 2 times faster in the best case, while on average estimating sensitivity accurately using 20\% of the full multiverse.
To enable analysts to stop early to fix errors or decide when results are ``good enough'' to move forward, we visualize both effect size and decision sensitivity estimates with confidence intervals, and surface potential issues including runtime warnings and model quality metrics.
}


\CCScatlist{ 
 \CCScat{K.6.1}{Management of Computing and Information Systems}%
{Project and People Management}{Life Cycle};
 \CCScat{K.7.m}{The Computing Profession}{Miscellaneous}{Ethics}
}

\figureTeaser

\vgtcinsertpkg

\begin{document}




\maketitle

\section{Introduction} 

Drawing inferences from data typically requires many analytic decisions:
Which data values are considered outliers? Which variables should be included as covariates? What model family and parameterization are appropriate?
At each decision point, more than one defensible option can exist.
When researchers explore different analytic paths but selectively report the ``best'' path leading to desired outcomes, they face the risk of inflated false-positive rates~\cite{simmons2011}.
This practice, colloquially known as ``p-hacking''~\cite{nelson2018}, is believed to be one cause for the replication crisis affecting various scientific fields~\cite{baker2016, gelman2013garden, gelman2014}.
To eliminate undisclosed flexibility in decision making, an increasingly popular approach is \textit{pre-registration}~\cite{cockburn2018}, where analysts commit all analytic choices to a verifiable registry before collecting any data.
However, even by committing to a single analytic path, the conclusion might still be less rigorous, as multiple justifiable paths might exist and different paths might produce diverging conclusions.
In crowdsourcing data analysis, well-intentioned experts still produced largely different analysis outcomes when independently analyzing the same dataset~\cite{silberzahn2018, schweinsberg2021, huntington2021}.

In response, in \textit{multiverse analysis} researchers attempt to evaluate all combinations of ``reasonable'' decisions, execute an end-to-end script for each combination, and interpret the results collectively~\cite{simonsohn2015, steegen2016, patel2015, young2017}.
With multiverse analyses, analysts can gauge whether their conclusions are robust to sometimes arbitrary decisions, and identify decisions that have a large impact on results.
By reporting the full range of possible outcomes, not just those that fit a particular hypothesis or narrative, the transparency of the study is also improved~\cite{rubin2017}.

However, a major pain point in conducting a multiverse analysis is its computation cost.
As the number of scripts increases exponentially with the number of decisions, the resulting multiverse often contains a large set of scripts, with the median size in practice being in the thousands~\cite{liu2020-paths}.
Combined with the need to evaluate each script end-to-end, a large collection of computationally demanding analyses (\eg Bayesian models) might take hours or even days to run.
Furthermore, multiverse analysis workflows can be \textit{iterative}.
It can be challenging to construct an \apriori ``reasonable'' decision space based solely on theoretical and methodological concerns, as certain validity issues only arise during runtime, such as model convergence and goodness of fit~\cite{liu2020boba}.
As a result, analysts might revise the multiverse specification according to what seems reasonable post-hoc, run the multiverse again, and repeat the process if necessary.
Being able to assess and revise multiverse results efficiently is thus important.

To address these challenges, we first investigate approximation algorithms based on sampling:
given a decision space, might we find a subset of universes that provide a good estimate of key results?
Since the primary goal of multiverse analysis is to gauge the robustness of outcomes to analytic decisions, we would like the sampling algorithms to accurately estimate \emph{decision sensitivity}: to what extent outcomes vary across different options within a decision.
We also ensure that the algorithms estimate the \emph{mean effect size} with low bias.
By estimating these two types of multiverse results quickly, the algorithms reduce the time until results are ``good enough'' to move forward with additional analysis.
Sampling is also compatible with parallelism (\eg running universes across multiple processes), a common method for reducing latency.
As a secondary contribution, we evaluate four methods for quantifying the sensitivity of individual decisions and recommend a metric based on the K-samples Anderson-Darling test~\cite{scholz1987}.
This metric updates the selection of sensitivity metrics in earlier work~\cite{liu2020boba}.

Next, we propose a monitoring dashboard for analysts to track progress, identify issues, and control execution of the multiverse -- leveraging sampling-based approximation algorithms under the hood.
To help analysts determine when the sampling has achieved reasonable performance, the interface displays sampling estimates in real time with confidence intervals.
More importantly, the dashboard enables analysts to reflect on the decision space structure, review intermediate results, and identify issues including runtime warnings and model diagnostics, before the multiverse evaluation completes.
When issues are identified early, users can stop the execution, refine the multiverse specification, and then resume, thus preventing wasted effort in running a problematic multiverse specification to completion.
\tim{broader implications?} 

\section{Related Work}

We introduce the background, terminology and related work on multiverse analysis.
Our work draws on techniques in approximate query processing and parameter space sampling.

\subsection{Multiverse Analysis}

To conduct a multiverse analysis, analysts start by outlining a space of reasonable analytic choices~\cite{simonsohn2015, steegen2016}.
Here we use \textit{decision} to refer to an element that varies across different analyses, and \textit{option} to refer to a possible value that a decision can take.
Decisions can be either discrete with categorical options, or continuous with a range of valid values.
As most existing multiverse analyses consist of purely discrete decisions, we focus on discrete decisions in this paper, though our sampling methods can be applied to discretized continuous decisions.

With a set of reasonable decisions, analysts then systematically enumerate all compatible combinations to create the decision space.
Analysts start with a cross-product of all decisions, then exclude invalid combinations, \vv{for example only Bayesian methods include decisions about priors.} 
We use the term \textit{procedural dependency} to denote the relationship between decisions that have invalid combinations.
Analysts then run the end-to-end analysis script for each decision combination to obtain a collection of results.
We call each script a \textit{universe}, and an output from the script a multiverse \textit{outcome}.

As the goal of a multiverse analysis is to gauge the range of results across all reasonable specifications, the multiverse should only consist of reasonable decision combinations.
Some prior work advocates that all reasonable decisions must be identified \apriori based on theoretical and statistical support~\cite{simonsohn2015}, though not all options might be equally defensible~\cite{del2021}.
A subsequent study finds that some \apriori reasonable decision combinations can produce models of insufficient quality, and recommends pruning the decision space prior to a final inferential step~\textit{post-hoc}~\cite{liu2020boba}.
We build on the latter idea to support early assessment of validity.

To interpret multiverse results, analysts first examine the range and distribution of all outcomes to gauge the overall robustness.
Visualizations support this task by displaying outcomes from individual universes using juxtaposition~\cite{simonsohn2015, rae2019}, superposition~\cite{liu2020boba}, or animation~\cite{dragicevic2019}.
Alternatively, charts visualize aggregated outcomes~\cite{patel2015, steegen2016, young2017}, such as a histogram of effect sizes~\cite{steegen2016}.
If the overall robustness check indicates conflicting conclusions, a natural follow-up task is to investigate which decisions are most consequential.
To support this task, prior studies use static visualizations including tables~\cite{steegen2016}, matrices~\cite{simonsohn2015}, and small multiples~\cite{poarch2019}, or interactive linked views between outcomes and the decision space~\cite{liu2020boba}.
We extract requirements for our approximation algorithms based on these prominent interpretation tasks.

Our work extends Boba~\cite{liu2020boba}, a system for authoring and interpretation of multiverse analyses.
With Boba, analysts author a multiverse specification using a \textit{domain specific language} (DSL) that models the decision space.
Then, a \textit{command line interface} (CLI) generates end-to-end scripts per-universe, runs the scripts, and collects outputs.
After all universes are executed, analysts can interpret the results using the Boba Visualizer.
Besides supporting the common interpretation tasks discussed above, the Boba Visualizer allows analysts to examine model quality and remove universes with poor quality before making a final inference.
However, the Boba Visualizer requires all universes to be executed and thus introduces long latencies before surfacing issues in model quality that might lead to a revision of the decision space.
Our work enables analysts to examine intermediate results and potential issues while the multiverse is still running, and incorporates approximation algorithms to make intermediate results meaningful.
In addition, Boba applies two methods for measuring decision sensitivity, which are based on the F-test and the Kolmogorov--Smirnov test, respectively.
We improve on the sensitivity metric by conducting an empirical evaluation on these two methods as well as two new methods.

While multiverse analysis is a recent statistical concept, the Visual Analytics community has made ample contribution on related problems including parameter space exploration~\cite{sedlmair2014}, model ensembles~\cite{wang2018}, multi-model steering~\cite{das2019beames}, and model comparison~\cite{cashman2019}.

\subsection{Approximate Query Processing}\label{sec:database}
Approximate query processing produces approximate answers to a query (\eg to a database management system) using a representative subset of the data.
The goal is to give a reasonable answer quickly, as large data sizes may produce long latency and impede subsequent work.
Two branches of prior work are closely related to interactive visualizations: online aggregation and optimistic visualization.

The main idea behind online aggregation~\cite{hellerstein1997} is the continuous refinement of approximate answers, rather than waiting for the system to finish with some \apriori error guarantee.
Users are provided with an interactive interface where they can view the answers immediately, watch the answers being gradually refined, and stop the query as they deem fit.
The method also provides confidence intervals alongside the query answer to convey uncertainty.
Our visual interface adopts the online aggregation paradigm; however, differences in the problem setting between a database query and a multiverse analysis prevent us from directly applying prior techniques.
A large database is often many orders of magnitude larger (\eg tens of millions of records) than a multiverse.
The cost of processing a tuple in a database is almost negligible, while a universe script can be time consuming to run.
Finally, online aggregation methods are designed primarily for \textit{aggregation} operations such as sum and average; while we are interested in the average of effect size, we also seek to understand the relationship between the decision space and the outcome.

Methods for sampling and computing confidence intervals are key components of online aggregation.
Prior work often derives confidence intervals from Hoeffding's inequality~\cite{hoeffding1994}, a powerful technique that makes no assumption on the data distribution except that the data is bounded.
But as the error bounds derived from Hoeffding's inequality are very relaxed, the technique is practical for large databases but less so for a typical multiverse analysis.
In terms of sampling strategies, uniform random sampling and round robin stratified sampling are the most widely-used (\eg~\cite{hellerstein1997, hou1988}).
Various sampling techniques that focus on specific data properties have been proposed~\cite{alabi2016, kim2015rapid}, for example IFocus quickly finds the correct ordering of bars in a bar chart~\cite{kim2015rapid}.
We similarly aim to preserve the correct ranking of decision sensitivity, but IFocus is based on Hoeffding's inequality and thus requires large data.

Optimistic visualization~\cite{moritz2017} is a different approach for presenting approximation results in an interface.
Instead of watching the result to improve gradually, users of optimistic visualization dive into analysis on a ``good enough'' snapshot of the data, then check their observations against the precise result later.
Our interface helps users decide when the snapshot is good enough to start further ``optimistic'' analysis.

\subsection{Sampling a Parameter Space}

We seek to understand the relation between decisions and multiverse results: what decisions lead to large variations in the results?
If we view the end-to-end analysis as a function and decisions as input parameters, we are interested in the relation between the parameters and the output.

\textit{Parameter space analysis} systematically varies the input parameters of simulation models to study the relationship between parameter combinations and corresponding outputs~\cite{sedlmair2014}.
The parameter space is continuous and thus needs to be coarsely sampled for a systematic exploration.
Common sampling methods include regular Cartesian sampling and uniform random sampling~\cite{sedlmair2014}. 
Prior studies also develop interactive tools for steering parameter space exploration, with the goal of finding interesting regions in the parameter space that lead to qualitatively different outputs~\cite{bergner2013, pajer2016}.
As a primary goal of multiverse analysis is an accurate characterization of the decision space, we apply sampling such that the overall decision sensitivity is estimated quickly.

%

Prior work in hyper-parameter optimization~\cite{bergstra2011, thornton2013} also proposes methods for efficiently exploring a parameter space.
The goal is to evaluate a small subset of promising hyper-parameter combinations in order to find the best-fitting model.
Similar to decisions in a multiverse analysis, hyper-parameters can be both continuous and discrete, and can have procedural dependencies.
Automated methods require a quantifiable objective to maximize, such as the classification accuracy~\cite{bergstra2011, thornton2013}.
Interactive systems for model selection allow a more ambiguous objective and incorporate domain knowledge from users~\cite{das2019beames,muhlbacher2017}.
As opposed to choosing an optimal model, the ``best'' parameter combination of a multiverse analysis is unclear. Instead, the goal is to understand the possible range of results across all reasonable parameters.

\section{Requirements}

We now outline our goals for approximation algorithms and design requirements for monitoring visualizations.

\subsection{Requirements: Approximation Algorithms}

\textbf{Online Approach:} We would like to adopt an online approach such that intermediate results can be displayed progressively in a user interface.
The approximation algorithms need to support continuous update and refinement of results.

\textbf{Accurate Estimation of Key Aspects:} We would like important aspects of the multiverse analysis to be approximated accurately.
Among prior works that report multiverse analysis results, the primary task is to understand the robustness of results across all reasonable specifications.
In particular, do certain decisions lead to large variations in outcomes? If so, which decisions are the most sensitive?
Thus, accurate estimation of \textbf{decision sensitivity} is an important goal.
In addition, the average of all end-to-end analysis outcomes (\eg effect size) suggest the overall direction and magnitude of effect.
Therefore, the algorithms should also estimate the \textbf{mean outcome} accurately.

\textbf{Confidence Intervals:} When viewing intermediate results, analysts should be able to gauge the uncertainty of the sensitivity or mean estimates.
The algorithms should produce confidence intervals (CI) that can help analysts decide when the results are ``good enough''.

\subsection{Requirements: Monitoring Dashboard}

We would like a monitoring dashboard for analysts to inspect the multiverse structure and intermediate outcomes without waiting for the entire multiverse to finish running.
With the dashboard, analysts should be able to diagnose issues early and stop the multiverse execution, in order to iterate on the multiverse specification.
In addition, users should be able to gauge when the sampling has achieved reasonable convergence, and dive into the next phase of the multiverse analysis workflow in an optimistic manner.

We identify the following tasks that analysts need to perform:
\begin{itemize}

	\item T1. \textbf{Review Decisions}: The tool should visualize the decision space for analysts to reflect on the multiverse structure, answering questions like ``are possible decisions missing'' and ``does a decision dominate the space''?
	\item T2. \textbf{Control Execution}: Analysts should be able to start, stop, and resume execution. Controlling execution is useful in various scenarios, including pausing to reclaim computational resources temporarily, or to sanity check errors and potential issues.
	\item T3. \textbf{Assess Progress}: Analysts should be able to track the progress of multiverse execution, in particular mean effect size and decision sensitivity over time with confidence intervals. Viewing progress allows analysts to gauge when the sampling results are stable enough to afford an optimistic analysis. 
	\item T4. \textbf{Find Issues}: Analysts should be able to observe potential issues and decide whether to stop early and refine the multiverse. Potential issues include an abnormal range of estimates, program errors and warning messages, and poor model quality.
	\item T5. \textbf{Identify Causes}: When issues arise, the tool should facilitate analysts in identifying the decisions that lead to the issues. Finding the contributing factors allows analysts to discard invalid options or narrow down the subset of universes to debug further.

\end{itemize}

\section{Sampling-Based Approximation Algorithms}

%
%

\subsection{Sampling Algorithms}

We investigate three approximation algorithms based on random sampling.
Sampling lends itself naturally to progressive display as universes are drawn one at a time from the multiverse.
We choose two algorithms that might have an advantage in estimating decision sensitivity quickly, and a third algorithm -- uniform sampling -- to serve as a baseline.

\fakeTitle{Notation}
A multiverse consists of $m$ decisions $\{A_i, i = 1 ... m\}$.
Each decision $A_i$ has $k_i$ discrete options, forming the set $\D_i = \{a_{il}, l = 1 ... k_i \}$.
The decision space is initially the product space $\Theta = \D_1 \times ... \times \D_m$, but some combinations may be invalid (\eg frequentist models and priors).
All valid combinations give rise to $n$ universes, $n \leq |\Theta|$.
All decisions in a given universe are summarized in vector $s_j$ with $s_j(A_i) \in \D_i$.
Executing the end-to-end script of the $j$-th universe produces an outcome value (\eg effect size estimate), denoted as $y_j \in \R$.

\fakeTitle{Sketching}
The first sampling method is a sketching algorithm in linear experimental design~\cite{raskutti2016}, with the goal of selecting a subset of input data points such that the underlying linear model is estimated accurately.
To apply the sketching algorithm, we first assume that a linear model describes the relationship between the input decision combination and the corresponding multiverse outcome value.
The model is most likely ``wrong'', but it is useful because a large model coefficient may indicate that the corresponding decision variable has a large impact on the outcome.
Since the decisions are categorical variables, we apply one-hot encoding enc$:s \rightarrow \mathbb{R}^d$ to obtain the feature vector enc$(s_i) = x_i$.
Then, we define a linear model between $x_i$ and $y_i$, with $\epsilon_i \sim N(0, \sigma^2)$ accounting for the model mismatch:

\begin{equation}\label{equation:linear-model}
\baselineskip=0pt
	y_i = x_i^T\beta + \epsilon_i \quad i = 1 ... n
\end{equation}

The sketching algorithm aims to find a small number of $(x_i, y_i)$ such that the estimated $\beta$ from the reduced version of the data is not too much different from the original Least Square estimator~\cite{raskutti2016}.
It relies on statistical leverage scores, where each leverage score corresponds to a feature vector $x_i$.
A higher leverage score indicates that the universe is more influential on the estimation.
If we stack $x_i$ as rows in matrix $X \in \R^{n \times d}$, the leverage score of the $i$-th universe is:

\begin{equation*}
	l_i = (UU^T)_{ii}
\end{equation*}

\noindent
where $X = U \Lambda V^T$ is the singular value decomposition for $X$.
\tim{optimality condition?} 
The algorithm then samples universes according to an independent distribution $(p_i)^n_{i=1}$, where $p_i = \frac{l_i}{d}$.

\fakeTitle{Round Robin}
The second algorithm is adapted from round robin stratified sampling, a widely-used sampling strategy in online aggregation of database systems (\autoref{sec:database}).
In each round, the algorithm goes over each decision and subsequently every option of the decision, and samples a universe uniformly at random from all remaining universes adopting the option.
This procedure ensures that for each decision option we will eventually have at least one multiverse covering that decision.
The pseudocode is shown in Algorithm~\ref{code:round-robin}.

\begin{algorithm}
\caption{Round Robin}\label{code:round-robin}
\begin{algorithmic}[1]
\State Initialize $T \leftarrow \varnothing$
\While{$ |T| < n$}
\For{each $A_i \in \{A_1, ..., A_m\}$}
  \For{each $a_j \in \D_i$}
    \State $S \leftarrow \{s_k  ~|~ s_k(A_i) = a_j ~\text{and}~ s_k \notin T \}$
    \If{$S \neq \varnothing$}
      \State Draw a sample $s$ uniformly at random from $S$
      \State $T \leftarrow T \cup \{s\}$
    \EndIf
  \EndFor
\EndFor
\EndWhile
\end{algorithmic}
\end{algorithm}
\vspace{-10pt}

\fakeTitle{Uniform Sampling}
As a baseline method, we also apply uniform sampling that draws each universe with equal probability.
Specifically, the algorithm samples universes from a discrete uniform distribution, where the probability of drawing the $i$-th universe is $p_i = \frac{1}{n}$.

\subsection{Correcting for Bias in Mean Estimation}\label{sec:correct-bias}

Intuitively, when the decision space is not a Cartesian product of all decisions, sketching and round robin algorithms over-sample certain regions of the decision space.
Sketching assigns higher leverage scores to data points that have a larger influence on the estimation of the linear model.
Round robin ensures that at least one sample is drawn from each option in a round, including rare options that would have a very low probability of being drawn in uniform sampling.
As a results, the sampling is biased.
This bias is intended, as some outcomes $y_i$ are important to know for gauging sensitivity, yet unlikely to be sampled.
We would like to correct for the bias such that we are not changing the final estimator of the mean outcome.

We apply \textit{importance sampling}, where the goal is to estimate properties of a distribution with only samples generated from a different, ``biased'' distribution.
Importance sampling corrects for the use of the biased distribution by applying weights given by the likelihood ratio:

\begin{equation}\label{equation:sample-mean}
	\bar{y} = \frac{1}{|T|} \sum_{i \in T} \frac{y_i f(y_i)}{g(y_i)}
\end{equation}

\noindent
where $f$ is the probability density function of the target distribution and $g$ is the probability density function of the biased distribution.
In our case, the target distribution is the discrete uniform distribution, $f(y_i) = \frac{1}{n}$.
The sampling distribution of the sketching algorithm, as described in the previous section, is $g(y_i) = \frac{l_i}{d}$.

We compute $g(y_i)$ of the round robin algorithm as follows.
Recall that in each round, the algorithm goes over every decision $A_j$, and a universe would have a matching option for each of the decisions.
Thus, the universe might be selected from any of the decisions, but the universe can appear in the sample only once.
With independent sampling between decisions, the inclusion probability for the $i$-th universe is

\[
	g(y_i) = \sum_{j} q_{ij} - \mathop{\sum\sum}_{j < k} q_{ij} q_{ik} + \mathop{\sum\sum\sum}_{j < k < l} q_{ij} q_{ik} q_{il} ...
\]

\noindent
where $q_{ij}$ is the probability that the $i$-th universe is selected from the $j$-th decision.
Basically, the equation calculates the probability that the $i$-th universe is selected by the first decision or the second decision or the third ...
As we sample uniformly from each stratum, $q_{ij}$ is simply the inverse of the size of the stratum that the universe belongs to.

\subsection{Quantifying Decision Sensitivity}\label{sec:sensitivity}

Because decision sensitivity is an important goal of our approximation algorithms, we need a method to quantify it.
Boba~\cite{liu2020boba} proposes two metrics, but in practice, we find that these methods can disagree.
To gain a deeper understanding of different sensitivity metrics, we conduct an evaluation on four candidate methods using synthetic datasets.
Here, we briefly introduce the methods and take-away results, and refer readers to the supplemental material for more details.


\figureResultBox
\subsubsection{Metrics}

We compare the following methods:

\begin{itemize}
\item The \textit{F-test} quantifies how much a decision shifts the mean of outcomes compared to the variance. If some options produce very different means than others, the decision may be highly sensitive. 
\item The \textit{Kolmogorov–Smirnov (K--S) test} is a non-parametric method to quantify the difference between two distributions. Using the test, we measure how different the outcome distributions are across different options of a decision. If certain options lead to very different distributions, the decision may be highly sensitive. 
\item The \textit{K-samples Anderson–Darling (AD) test} is another non-parametric method for establishing differences in two or more distributions~\cite{scholz1987}. Intuitively, the AD test is similar to the K--S test, but it compares $k$ options simultaneously.  
\item \textit{Linear regression (LR)} models a multiverse outcome as a linear combination of decision values (\autoref{equation:linear-model}). We would expect highly sensitive decisions to be associated with large coefficients. 
\end{itemize}

\subsubsection{Summary of Sensitivity Evaluation}


\textbf{False Positives:} Could a decision appear more sensitive simply because it has more options?
We construct a synthetic multiverse with only a non-sensitive decision (using the same general scheme as in \autoref{sec:dataset}), then vary the number of options within this decision.
We find that the sensitivity score of this non-sensitive decision in K--S test and LR increase with decision cardinality.
In other words, K--S test and LR may consider a decision to be more sensitive simply because the decision has more options.

\textbf{Non-Normality:} We then check how well the sensitivity tests handle non-normal distributions. 
We simulate decisions where the options are from different distributions (\eg normal versus Poisson), but have the same mean and standard deviations.
Being parametric methods, F-test and LR cannot reliably distinguish these different distribution functions. For example, a normally-distributed option and another lognormal-distributed option appear the same to F-test and LR.

\textbf{Sensitivity Ranking:} Finally, we systematically construct a series of synthetic multiverses with multiple decisions, then assess how well the metrics estimate the correct ranking of sensitive decisions.
F-test and AD test correctly recover the ranking in nearly all datasets, while K--S test and LR fail to do so in over half of the datasets.

Based on these take-aways with supporting evidence in Supplemental Section 1, we find the AD test to be the most effective for quantifying sensitivity. We use it in all subsequent experiments and recommend its use over the F-test and K--S metrics previously supported by Boba~\cite{liu2020boba}.

\subsection{Confidence Intervals}

We must calculate 95\% confidence intervals (CI) for both decision sensitivity and outcome mean estimates.
A widely-used method in database online aggregation literature is using Hoeffding's inequality to approximate the size of the confidence intervals (\autoref{sec:database}), but we find the CIs to be too loose for the common size of multiverse analysis.
Instead, we use bootstrapping: sampling with replacement within the universes drawn so far, we compute the decision sensitivity or bias-corrected mean to construct a resampling distribution, and obtain a CI from this resampling distribution.
To account for potential skew in the resampling distribution, we use the \textbf{bias-corrected and accelerated bootstrap}~\cite{efron1987} to derive CIs.

\section{Evaluation of Approximation Algorithms}

We perform empirical validation of the sampling algorithms in a suite of synthetic and real multiverse analyses.
We first measure how quickly the algorithms \textit{approximate} and \textit{rank} sensitive decisions.
Then, we evaluate the method for correcting bias in mean estimation.

\subsection{Datasets}\label{sec:dataset}

We design five synthetic datasets that are inspired by characteristics of real multiverse analyses.
Building these synthetic datasets allow us to control the data generating process and tease apart interesting properties of the multiverse to study each property in isolation.
We also run the benchmark on a multiverse analysis in the wild~\cite{schweinsberg2021} to demonstrate the utility of the sampling algorithms in real-world scenarios.

We now describe the general scheme for constructing the synthetic dataset.
We would like the synthetic data to contain both signal and noise to better simulate real-world scenarios, yet we need to have control over how much a decision influences the outcome.
Thus, we model each option $a_i$ within a decision as a normal random variable $N(\mu_{a_i}, \sigma^2)$, with a larger difference in $\mu$ between options indicating a more sensitive decision.
The outcome of a universe is then the sum over the contributions from all decisions.
Specifically, given an input vector of the $i$-th universe $s_i$, the outcome is

\[
	y_i = \sum_{j=1}^{m} Z_j, \quad Z_j \sim N(\mu_{s_i(A_j)}, \sigma^2)
\]

\noindent
Multiverses may contain decisions that are not important at all (\eg \cite{young2017}).
To model these, we first define a \textit{baseline option} by setting the mean to zero, such that the option contributes nothing but random noise to the outcome.
We then construct \textit{non-sensitive decisions} by setting every option to a baseline option. 
Multiverses may also contain certain \textit{rare} conditions -- the number of universes adopting a particular option is smaller compared to the number of universes adopting other options (\eg \cite{schweinsberg2021}).
We capture these by simulating procedural dependencies, which exclude invalid combinations from the Cartesian product decision space.
In building synthetic datasets, we also seek to make the total size, the number of decisions, and the cardinality of decisions as realistic as possible~\cite{liu2020-paths}.
The sizes of the synthetic multiverses range from 200 to 1,552, with 4 to 8 decisions and 2 to 10 options per decision, unless stated otherwise.
We now introduce the characteristics of each dataset.

\fakeTitle{D1: Simplest}
We design this synthetic multiverse to be the simplest scenario where we would expect the sampling algorithms to perform differently.
The multiverse contains four non-sensitive decisions and one sensitive decision with some baseline zero-mean options, as well as an influential option with a large mean $\mu_{a_i} = 6 \sigma$.
This influential option is relatively rare due to invalid combinations.
We expect both sketching and round robin to have a different probability of including the influential option compared to uniform sampling.

\fakeTitle{D2: Interaction}
Decisions may interact in real-world multiverses, where the impact of one decision on the outcome depends on the option chosen another decision.
We construct a synthetic multiverse with eight binary decisions, six of which are non-sensitive.
The remaining two sensitive decisions interact, and one of the interacting combination is relatively rare.

\fakeTitle{D3: High Cardinality}
This dataset aims to evaluate how well sampling algorithms perform on a decision with a large number of options.
The multiverse has one sensitive decision with 50 options, including 45 baseline options and 5 options with non-zero means.
Half of the options, including the 5 impactful options, are relatively rare.
The other decisions are non-sensitive.

\fakeTitle{D4: Distractor Decisions}
Building upon the simplest scenario D1, we would like to ``distract'' the sampling algorithms in finding the rare, sensitive decision by including other sensitive decisions.
The multiverse has three sensitive decisions and four non-sensitive ones.
Similar to D1, a sensitive decision consists of a rare influential option among other baseline options.
The other two sensitive decisions do not involve procedural dependencies, serving as the ``distractors''.

\fakeTitle{D5: Distractor Options}
Building again on the simplest scenario D1, we now seek to distract the sampling algorithms by making other options rare.
The synthetic multiverse has the same composition as D1, except that four baseline options of the sensitive decisions are relatively rare, in addition to the influential option.

\figureResultPearson
\figureResultSpearman
\fakeTitle{D6: Real Multiverse}
The last dataset is a real-world multiverse analysis, taken from a study on crowdsourced data analysis~\cite{schweinsberg2021}.
The multiverse has seven decisions, including different ways to operationalize dependent and independent variables, choose the model family, and pick the set of covariates.
We choose this dataset because the decisions have complex dependencies and the multiverse is large in size with 2,977 universes.

\subsection{Evaluation: Decision Sensitivity}

Using the six datasets, we first evaluate how well the sampling algorithms estimate decision sensitivity.

\subsubsection{Methods}
We assess how quickly the algorithms \textit{approximate} and \textit{rank} sensitive decisions.
To this end, we define a set of termination conditions, sample universes progressively, and record the percentage of the full multiverse drawn when the termination conditions are met.
The termination conditions consist of three requirements.

First, the estimated sensitivity must closely match the ground truth.
We quantify the sensitivity of individual decisions using the standardized test statistics of the k-samples AD test, which produces a list of sensitivity scores.
We then compute the Pearson correlation of the sensitivity scores between the sample and the full data.  
The first condition requires the Pearson correlation coefficient to be larger than 0.95.

Second, the sensitive decisions must be ranked correctly.
Here, we discard non-sensitive decisions, because non-sensitive decisions do not have a clear ranking, yet may have slightly different sensitivity scores due to noise in the data generation process.
To assess ranking, we calculate the Spearman correlation between the sensitivity scores of sensitive decisions in the sample and the full data. 
The Spearman correlation must be 1 for the second condition to be met.

Third, each option must have at least three samples. This condition is a prerequisite for the k-samples AD test to work reasonably well~\cite{scholz1987}.

Because the samples drawn by all three sampling algorithms vary due to randomness, we repeat the benchmark 200 times using different random seeds and report the median performance.

While the method above summarizes performance into a single number, it relies on several cutoff values.
To further characterize performance, we examine Pearson and Spearman correlation (if applicable) as more samples are drawn. 
We average the correlations across 200 runs.
The correlation is null when the options do not have sufficient sample size for the AD test.
Because a sample producing a null correlation is useful information, we impute these null values using 0, as if there is no relationship between the sample sensitivity and the ground truth.

\subsubsection{Results}

\autoref{fig:result-box} shows the mean, median, and IQR of the proportion of the full multiverse drawn until the three termination conditions are met, across 200 runs.
In the simplest scenario D1, the sketching and round robin algorithms take 9.4\% and 9.5\% on average to meet the termination goals, respectively.
Both algorithms are over 2 times faster than uniform sampling (mean=20.2\%).
Compared to no approximation at all, using an approximation algorithm, even the baseline, is 5 times faster.

In two of the more complex scenarios, round robin outperforms the other algorithms by at least 2 times, while sketching is slightly better than the baseline.
On a multiverse with a high cardinality decision (D3), round robin is relatively fast (mean=18.4\%) in including all options needed for an accurate estimation, due to the structure imposed in the sampling procedure.
On the contrary, the random nature takes sketching (mean=41.6\%) and uniform (mean=48.0\%) much longer to gather sufficient samples across all options.
When distractions from other rare options are present (D5), round robin maintains a similar performance to D1 (mean=7.7\%), but sketching is considerably worse (mean=15.5\%).
In the remaining two synthetic datasets, sketching and round robin do not provide a performance gain over the baseline.

In D2, sketching offers no advantage in detecting interactions between decisions, since its underlying linear model does not include interaction terms.
As a follow-up exploratory analysis, we add all possible two-way interactions in the linear model, and sketching improves slightly by using 3\% fewer samples.

In D4, other sensitive decisions might indeed distract sketching and round robin from correctly ranking sensitivity.
Let $A_1$ be the rare sensitive decision and $A_2$ be the distractor decision.
In the full multiverse, $A_1$ is less influential than $A_2$.
As both algorithms over-sample $A_1$, the larger sample size with more ``evidence'' of impact leads both algorithms to believe $A_1$ to be more sensitive.
\autoref{fig:result-spearman}-D4 shows how the ranking of sensitive decisions is disrupted along the course, while overall sensitivity scores of all decisions are already very close to the full data (\autoref{fig:result-pearson}-D4).

Finally, both sketching (mean=10.4\%) and round robin (mean=9.7\%) are considerably faster on the real multiverse compared to the baseline (mean=16.2\%).
The trends of decision similarity (\autoref{fig:result-pearson}-D6) and ranking (\autoref{fig:result-spearman}-D6) over time show that round robin outperforms sketching, while sketching outperforms uniform sampling.

\subsection{Evaluation: Bias Correction in Mean Estimation}

We also evaluate the methods for correcting bias in the estimation of mean multiverse outcome, as described in \autoref{sec:correct-bias}.
Recall that uniform sampling gives an unbiased estimation of the mean, whereas sketching and round robin may shift the outcome by oversampling certain regions of the decision space.
The empirical evaluation seeks to validate that the two algorithms produce a mean estimate as accurate as uniform sampling, after bias correction.

In the experiment, we draw samples progressively until all universes are included.
We then calculate the mean squared error between estimated and actual mean:

\[
	\text{MSE}=\frac{1}{n-b}\sum_{i=b}^{n} (\bar{y_i} - \mu)^2
\]

\noindent
where $\bar{y_i}$ is the estimated mean (\autoref{equation:sample-mean}), $\mu$ is the true mean, and $b$ is a parameter for discarding large errors in very small samples.
We set $b$ to be the sum of the cardinality of all decisions.
To offset randomness, we repeat the experiment 200 times using different random seeds.

\autoref{fig:result-bias} shows the MSE of 200 runs on D4.
The result confirms that the arithmetic mean of sketching and round robin does differ considerably from the actual mean, but the bias correction method successfully brings MSE to a value similar to uniform sampling.
Results from other datasets show a similar pattern and are supplemental material.
\tim{summary paragraph?} 

\figureResultBias

\section{Progressive Visualization: The Boba Monitor}

Next, we introduce the visual design and interactions of the Boba Monitor, an extension to the Boba system.
The Boba Monitor is a dashboard for users to control and monitor multiverse execution.
It serves as a bridge between authoring the multiverse specification using the Boba DSL, and analyzing the multiverse results using the Boba Visualizer.
Before opening the dashboard, users first need to author the multiverse specification, including the shared portion of the analysis script and corresponding decisions.
After using the dashboard, users either go back to iterate on the multiverse specification, or proceed to the next stage of multiverse analysis using the Visualizer.

\subsection{Monitoring Progress}
The upper part of the dashboard is primarily for controlling the execution and observing the progress as the multiverse continues to run.

\subsubsection{Control Panel}

The control panel (\autoref{fig:teaser}a) allows users to directly control the multiverse runtime from within the dashboard.
It supports starting, stopping and resuming the execution (T2).
The panel also displays a progress bar to indicate the proportion of universes completed so far, along with an estimated completion time.
The progress bar and estimated completion time are updated in real time as the multiverse runs in the background, allowing users to continuously track the basic status of execution (T3).
From a user experience perspective, a progress indicator alone can improve the perceived speed of the system~\cite{myers1985}.

Before invoking the multiverse runtime, the system applies one of the approximation algorithms to calculate the sampling order and executes universe scripts according to the order.
By default,  we use the round robin algorithm, but users can override the default in configurations.
As opposed to running the multiverse sequentially, apply a sampling algorithm is crucial because it not only reduces the time until decision sensitivity is estimated accurately, but also ensures that the intermediate results are not misleading.

\subsubsection{Progress View}

The progress view (\autoref{fig:teaser}b) displays the trends of sampling estimates over time, enabling users to continuously observe progress and determine when the estimates are good enough (T3).
The view includes two line charts: one visualizes the time-series of the average multiverse outcome, and the other show a collection of time-series of decision sensitivity score, one per decision.
Around each line, an area band indicates the 95\% confidence interval of the corresponding estimate.
Similar to the control panel, these charts are updated continuously, but with a lower frequency as bootstrapped confidence intervals can be costly to compute.

The primary purpose of this view is giving users a sense about when to dive in optimistically to the next analytic phase.
Users might decide that results are ``good enough'' to proceed when the mean effect size chart shows a stable trend with a narrow confidence interval that does not overlap with the critical value (\eg zero).
Similarly, in the decision sensitivity chart, the sensitivity scores of important decisions might need to be well-separated, with non-overlapping confidence intervals.

\subsection{Diagnosing Issues}
The bottom part of the dashboard consists of multiple coordinated views to facilitate users in identifying potential issues in the multiverse.

\subsubsection{Decision Space View}

The decision space view (\autoref{fig:teaser}c) visualizes the decisions and their relationship to support reflection on the decision space structure (T1) and to contextualize subsequent tasks.
The visualization is present before any universes are run, giving users a chance to review the decisions and potentially go back to revise the multiverse specification early on.

We visualize the decision space using Analytic Decision Graphs~\cite{liu2020-paths}, a design previously adopted by the Boba Visualizer, to maintain consistency across system components. 
In the graph, nodes represent decisions, with a larger size indicating a higher cardinality and a darker color indicating a higher sensitivity.
Edges depict relationships between decisions in the analysis process, including  order (light gray edge) and procedural dependency (black arrow).
Options of a decision are shown besides the decision node.
The underlying data for the graph is extracted automatically from the multiverse specification written by the author in the Boba DSL.
Decision sensitivity is not available before runtime, but is gradually updated as the multiverse executes.

\figureModelQuality
\subsubsection{Main Effect View}
The main effect view (\autoref{fig:teaser}d) visualizes outcomes from individual universes as a dot plot, with one-to-one mapping between a dot and a universe.
The view serves three main purposes.
First, it conveys the range of possible outcomes for users to spot potential issues of abnormal outcome values (T4).
Second, it supports an overview of other type of issues, including errors and model quality, by encoding the corresponding aspect of a universe using color (T4).
Third, brushing and linking interactions with other views enable users to connect decisions to a particular issue and gain a better understanding of the responsible decisions (T6).

In the dot plot, the x-axis represents the magnitude of the outcome value and the y-axis represents count.
Outlying point estimates might be observed at the left and right extremes of the x-axis.
To bring attention to universes that do not produce an outcome (often due to errors causing the execution to fail), we show them in a special bin at the left end of the x-axis.
The color scheme of the dots encodes either error information or model quality, and can be toggled in a dropdown menu (\autoref{fig:system-model-quality}c).
A categorical scheme (\autoref{fig:teaser}d) encodes error information, indicating if a universe stopped with errors (red), completed with warnings (yellow), or completed without warnings (blue).
Alternatively, a sequential color scheme (\autoref{fig:system-model-quality}a) maps to a quantitative model quality metric, such as R-squared, with a lighter blue indicating poorer model quality.
The color encoding gives an overview of the issue and invites direct manipulation interactions to inspect details.

This view supports two types of brushing and linking interactions with other views.
First, brushing the main effects view filters the diagnostics to the selected universes.
The error message view and the model quality view are updated accordingly to detailed diagnostics restricted to the selected results.
Second, clicking a decision node in the decision view (\autoref{fig:teaser}c) splits the main effects into a small multiples plot, with each subplot mapping to a different option of the decision (\autoref{fig:teaser}d).
The small multiples enable analysts to compare between options and identify options that potentially lead to a particular issue.
For example, an option might exclusively produce outlying main effect estimates, runtime errors, or poor-fitting models.
Finding the responsible decision gives users actionable information to improve the multiverse.
Users might realize that certain options are unreasonable and exclude them from the multiverse, or otherwise narrow the issue down to a smaller subset of the multiverse for further diagnosis.

\figureMortgage
\figureCaseTwo
\subsubsection{Error Message View}\label{sec:error-view}
The error message view (\autoref{fig:teaser}e) compiles a list of distinct error and warning messages from the runtime to give users more details on implementation issues (T4).
The messages are aggregated since the same runtime error often gets repeated across multiple universe scripts.
This is because universe scripts are highly similar, for instance two adjacent universes that differ in only one decision may share the same source code except for the value of a variable.

To extract the error messages, the system first collects the outputs of universe scripts in the standard error stream. 
These outputs are then simplified and aggregated using heuristics that look for specific keywords such as \textit{warning}.
The heuristics take advantage of the shared logging format of a particular programming language, for example a Python program typically prints a stack trace that starts with the word \textit{Traceback}.
We find the heuristics work well in practice, though a more generalizable method could apply sentence embeddings~\cite{cer2018} to compute the similarity between error messages and look for repeated messages.

In the interface, we assign a categorical color scheme according to the type of the message: red for errors that terminate the program and produce a non-zero exit code, and yellow for warnings.
The messages are ordered first by their type (\ie errors before warnings) and then by the number of universes that share the issue, to prioritize more important messages at the top.
As described before, brushing the main effect view filters messages to be those that occur in the selected universes.
In addition, clicking a message updates the main effect view to highlight universes with that specific message.

\subsubsection{Model Quality View}\label{sec:fit-view}
The model quality view (\autoref{fig:system-model-quality}b) supports an in-depth assessment of the model quality of individual universes using visual predictive checks (T4).
A dropdown menu (\autoref{fig:system-model-quality}c) toggles between the model quality view and the error message view.
The predictive check allows users to detect systematic discrepancies between the model and observed data in order to assess the fit of the model to the data~\cite{gelman2006book}. 
It compares the replicated dataset from predictions of the fitted model to the observed dataset.
We facilitate the qualitative comparison by plotting the predicted and observed data distributions side-by-side as two violin plots.
We also overlay a representative subset of individual data points to convey further details.
This subset is derived by sampling in each percentile of the corresponding distribution and visualized as a centered quantile dot plot~\cite{kay2016when}.

\section{Case Studies}

\vv{We evaluate the Boba Monitor via two case studies to demonstrate how it helps users in gauging sensitivity and diagnosing issues early.}

\subsection{Case Study: Mortgage Analysis}

In the first case study, we demonstrate that by running a small subset of a multiverse, we can arrive at the same conclusion about decision sensitivity and model quality as prior work.

Young~\etal develop a method for multimodel analysis~\cite{young2017}, with the goal of assessing to what extent model results are robust to changes in model specification.
The method collects model results across all combinations of sensible control variables, then quantifies how much a variable influences the coefficient of interest.
Using our terminology in this paper, the method runs a multiverse where each decision is whether to include a control variable, and quantifies decision sensitivity.

The original paper~\cite{young2017} describes an example multiverse analysis.
Using publicly available loan-level data, the analysis aims to answer ``are female applicants more likely to be approved for a mortgage?''
To this end, the analysis uses a linear regression model with mortgage application approval as the dependent variable and a dummy variable indicating if the applicant is female as an independent variable.
To build the multiverse analysis, the authors take all combinations of reasonable control variables, such as loan history and demographic information.
These combinations give rise to 256 end-to-end analyses.
Based on the model coefficients of \textit{female} across all analyses, the method concludes that two control variables, \textit{black} and \textit{married}, are particularly sensitive, while the other variables are not.
Liu~\etal~\cite{liu2020boba} repeat the multiverse analysis and note model quality issues, as there is a large discrepancy between observed data and model predictions.

We load the same multiverse specification into the Boba Monitor.
After starting the runtime, we observe the change in the estimates of decision sensitivity and mean effect size, and pause the runtime when these estimates seem to reasonably converge.
\autoref{fig:mortgage} shows the snapshot when we pause, with 46 universes completed so far.
The mean effect size estimate (\autoref{fig:mortgage}b) fluctuates slightly at the beginning, but soon remains stable around the value of 2.2, with a reasonably narrow confidence interval spanning roughly 1.6 to 2.6.
The decision sensitivity estimates (\autoref{fig:mortgage}a) show that two decisions are sensitive while others are not, and the confidence intervals of the sensitive decisions do not overlap with those of the non-sensitive decisions.
The decision graph confirms that the sensitive decisions are \textit{black} and \textit{married} (\autoref{fig:mortgage}c), which agree with prior work.
We may also identify the sensitive decisions by hovering over the time series in the progress view.
This exercise shows that an optimistic analysis at 18\% of the full multiverse size will not lead to a qualitatively different interpretation of decision sensitivity than running every universe.

Before proceeding to further analysis and running the multiverse to completion, we sanity-check possible validity issues.
The main effect view shows that all model coefficients are within a sensible range.
The error message view shows that no runtime errors or warnings have occurred so far.
We then switch to examine the model quality, which we quantify using adjusted $R^2$ as all universes use a linear model.
As indicated by the light color (\autoref{fig:mortgage}d), the model quality of all completed analyses is unanimously bad.
The visual predictive checks (\autoref{fig:mortgage}e) show that the observed data is binary while the model prediction is continuous, suggesting that we need to revise the multiverse specification with a more appropriate model such as logistic regression.
Importantly, we are able to identify the model quality issue before one fifth of the multiverse finished running (or as early as the first few universes, if we check earlier), reducing the time to iteration.

\subsection{Case Study: Gender in Scholarly Debates}

The second case study illustrates in more detail how the Boba Monitor facilitates early error diagnostics.
To capture issues that may arise from the early stage of developing a multiverse analysis, we build a multiverse from scratch using decisions made by crowdsourced analysts on a shared dataset~\cite{schweinsberg2021}.
The dataset contains words and metadata of scientific debates from an invitation-only online forum for scholars.
Different teams of analysts are tasked with analyzing the data to answer ``does womens' tendency to participate actively in a conversation increase with more woman participants?''
Analysts make different choices in operationalizing the dependent and independent variable, transforming the data, choosing covariates, and specifying the statistical model.
We would like to build a multiverse from these analytic decisions.
In practice, even with an \apriori decision space, it is usually easiest to start with a simple multiverse and then build in additional complexity, checking for issues along the way.
To simulate the process, we start with a subset of the decisions and take all compatible combinations between them, which results in 468 universes.

After writing the multiverse specification, we run the multiverse in the Boba Monitor.
We observe plenty of warnings while the sampling estimates have not yet converged, and pause the runtime to investigate.
\autoref{fig:case} shows the snapshot of the multiverse when 79 universes are completed.
We first look for potential issues from the overview in the main effect view.
The point estimate of this multiverse is the \textit{z-score}, which measures how many standard deviations away the raw estimate is from the mean.
From the range of z-scores, we can see disproportionate estimates that are more than a hundred standard deviations from the mean.
In addition, almost half of the universes output an error or warning message.
Switching to viewing the model quality, we observe that many models are of unsatisfactory quality.
These observations suggest a need to revise the multiverse specification.

To gain a better understanding of the directions of improvement, we gauge the validity of decisions by clicking on the decision nodes and comparing the options.
As shown in \autoref{fig:case}a, the Poisson regression model is responsible for all errors,  warnings, and abnormal z-score estimates.
\autoref{fig:case}b shows that adding random effects greatly improves model quality and eliminates outlying point estimates, which suggests that a single-level regression model is probably insufficient to describe the data.
We might consider removing the option of single-level models from the decision space in subsequent iterations.

Next, we take a closer look at the runtime errors and warnings in the error message view.
We first click on the \textit{show more} button to view each message in full.
Some errors are programming oversights such as a null reference error, requiring a simple fix.
Some warnings are related to model convergence and might be improved by increasing the number of possible iterations or switching to a different optimizer.
However, fixes for other issues may require knowing the accountable decisions.
For example, the warning \q{some predictor variables are on very different scales} indicates problems with predictors, yet the multiverse has many combinations of independent and control variables.
We click on the message to highlight related universes in the main effect view, and click the \textit{IV} and \textit{Covariates} decision nodes.
The warning occurs in both independent variable options, but only appears in one set of covariates (\autoref{fig:case}c), which suggests that rescaling the variables in this set of covariates should fix the issue.

\section{Conclusion and Future Work}

This paper presents methods for reducing the delay from having a multiverse specification to checking intermediate results and potential issues.
We adopt an online approach and investigate three sampling-based approximation algorithms.
\vv{Empirical evaluation on synthetic and real multiverses show that round robin and sketching approaches are 5 times faster on average to estimate decision sensitivity accurately compared to no sampling, and up to 2 times faster than uniform sampling.}
Next, we develop a monitoring dashboard that leverages sampling algorithms under the hood.
In two case studies, we demonstrate how the interface facilitates users in choosing a ``good enough'' snapshot for early interpretation of multiverse results, as well as diagnosing runtime errors and model quality issues.
We now discuss limitations and future work.

\fakeTitle{Adaptive Sampling}
In this work, we investigate sampling algorithms that do not depend on the multiverse outcome, but instead rely solely on the structure of the input decision space.
Alternatively, an adaptive sampling approach might select new samples based on knowledge of previous samples, for instance sampling more heavily from decisions that seem to exhibit high sensitivity so far.
We do not adopt the later approach because we would like an accurate characterization of the decision space as a whole.
However, with continuous decisions, users may wish to obtain better resolution on sensitive regions after an initial characterization of the full space, for which adaptive sampling is effective.
Future work might investigate interactive strategies for steering the exploration~\cite{bergner2013, pajer2016} of a continuous decision space of the multiverse.

\fakeTitle{Reducing Latency}
To reduce delay in viewing multiverse results, we apply approximation algorithms, but this does not preclude other methods in latency reduction. 
A simple extension is parallel computing, with universes executed on separate processes simultaneously.
Sampling is amenable to parallel computing as along as the sampling order is preserved.
More involved techniques might attempt to optimize the runtime directly, as universe scripts are often highly similar with a majority of redundant source code.
Future work could investigate dynamic program analysis and caching to reduce repeated computation.

The Boba Monitor largely adopts an online approach to display sampling estimates and confidence intervals progressively, and relies on analysts to decide when to view optimistic visualizations on a ``good enough'' snapshot. 
We imagine a system that automates this decision, for example recommending a snapshot when the ranking of decision sensitivity is likely to be correct with high probability~\cite{kim2015rapid}.
The system can also do more to communicate the discrepancies between interpretations of optimistic visualizations and the ultimate precise results.

\fakeTitle{Target Users}
The intended users for our system are \textit{authors} of the multiverse analysis, who we assume would have a high level of statistical expertise in constructing, diagnosing, and interpreting the analyses.
There remains important future work to further lower the barriers for less experienced analysts.
One direction is to present users with high-level abstractions that represent expert statistical knowledge, and synthesize appropriate specifications from high-level analysis goals~\cite{jun2019}.
Another is to guide users in defining the decision space, for example restricting the set of reasonable decisions based on analytic assumptions and data characteristics~\cite{jun2019}, or expanding the space by recommending additional decisions. 
\vv{In addition, a multiverse analysis shows the uncertainty in decision variations, but people have difficulties in interpreting uncertainty representations~\cite{hullman2019}.
How to communicate multiverse results to a more general audience is an open question that benefits from the ongoing research on effective uncertainty visualizations.}

\fakeTitle{Model Expansion}
In practical data analyses, it is recommended to start with a simple model, perform checks and diagnostics once the model is fit, and expand the model with further complexity incrementally~\cite{gelman2006book}.
By visualizing posterior predictive checks of a partial multiverse, users of our system might adopt a similar workflow to improve a multiverse based on model diagnostics. 
 However, it is unclear how to integrate the model expansion and multiverse analysis workflow: does the analyst incrementally build a series of models and construct the multiverse out of all reasonable ones, or start with an \apriori decision space and prune away models that fail the diagnostics, or both?
 Furthermore, while posterior predictive checks suggest directions to improve a model, it is difficult to do so across many different models.
 As a starting point, future tool might automatically cluster similar patterns in the predictive checks such that analysts might formulate refinement strategies for a set of universes instead of one at a time.

\acknowledgments{This work was supported by NSF Award 1901386.}



\end{document}